\documentclass[
amsmath,amssymb,
aps,
reprint,
nofootinbib,
]{revtex4-2}

\usepackage{graphicx}
\usepackage{dcolumn}
\usepackage{bm}
\usepackage{color}
\usepackage{hyperref}

\newcommand{\bea}{\begin{eqnarray}}
\newcommand{\eea}{\end{eqnarray}}
\newcommand{\be}{\begin{equation}}
\newcommand{\ee}{\end{equation}}
\newcommand{\ba}{\begin{align}}
\newcommand{\ea}{\end{align}}

\begin{document}

\title{
Shear mode transport coefficients from multiple polylogarithms
}

\author{Paolo Arnaudo}
 \email{p.arnaudo@soton.ac.uk}

\affiliation{Mathematical Sciences and STAG Research Centre,\\ University of Southampton, Highfield,\\ Southampton SO17 1BJ, UK}%

\begin{abstract}
We present an analytical study of the transport coefficients associated with the shear sector of gravitational perturbations around asymptotically anti-de Sitter black branes.
In the long-wavelength, low-frequency limit, the wave solutions admit a structure that is fully described in terms of multiple polylogarithms in several variables.
We focus primarily on computing the transport coefficients for $\mathcal{N}=4$ SYM, by performing a bulk computation in the five-dimensional black hole background up to order $\mathfrak{q}^{10}$, which extends the results previously available in the literature.
We then generalise the procedure to $d+1$ dimensions, characterising the mathematical structure of the resulting transport coefficient expressions.
\end{abstract}

\maketitle


\section{Introduction}

The computation of transport coefficients via holographic duality provides a crucial window into the non-equilibrium dynamics of strongly coupled quantum field theories. In the framework of the AdS/CFT correspondence, gravitational perturbations around black brane or black hole backgrounds encode the dissipative and hydrodynamic response of the boundary theory \cite{Son:2002sd, Policastro:2002se, Policastro:2002tn, Kovtun:2005ev}. 
In the long-wavelength, low-frequency limit, the linear response of the boundary stress–energy tensor to boundary metric perturbations defines a set of transport coefficients that characterise the system’s approach to equilibrium. Beyond the leading-order terms, one encounters higher-order corrections that reveal additional dissipative effects and the causal structure of the system \cite{Withers:2018srf, Grozdanov:2019kge, Grozdanov:2019uhi, Heller:2020uuy, Heller:2020hnq, Heller:2022ejw, Heller:2023jtd}. These higher-order coefficients can be systematically extracted from subleading terms in the perturbative bulk solution. 
The knowledge of transport coefficients in holography has historically provided notable and influential results in the context of the quark-gluon plasma, and more generally within the broader AdS/CMT program \cite{Policastro:2001yc, Kovtun:2004de, Hartnoll:2009sz, AdSCMTbook}.

The goal of this paper is to study the analytic structure of the transport coefficients for the shear sector of gravitational perturbations around the planar $d+1$-dimensional Schwarzschild anti-de Sitter (SAdS$_{d+1}$) backgrounds. The master equations for these perturbations were analysed in \cite{Kodama:2003jz}.
More precisely, denoting with $R$ the radius of AdS, the planar SAdS$_{d+1}$ metric reads\footnote{Comparing to the notations in \cite{Kodama:2003jz}, we are taking $K=0$, $\lambda=-1/R^2$, $M=\mu/(2R^2)$.}
\begin{equation}\label{metricSAdS}
\begin{aligned}
&\mathrm{d}s^2_{d+1}=-f(r)\,\mathrm{d}t^2+\frac{\mathrm{d}r^2}{f(r)}+\frac{r^2}{R^2}\sum_{i=1}^{d-1}\mathrm{d}x_{i}^2,\\
&f(r)=\frac{r^2}{R^2}-\frac{\mu}{R^2\,r^{d-2}},
\end{aligned}
\end{equation}
where $\mu$ (having the dimension of the $d$th power of a length) parametrizes the mass of the black hole. In all the computations, we will fix the normalisation $\mu=R=1$.

We start by showing the results in the extensively studied $d=4$ case, which corresponds to the D3-brane background, and, via AdS/CFT duality, to $\mathcal{N} = 4$ super Yang-Mills (SYM). 
We obtain in particular the analytic expression of one additional transport coefficient beyond those previously known in the literature (see for example \cite{Natsuume:2007ty, Grozdanov:2019uhi}).

We then generalise the construction to the generic $(d+1)$-dimensional case, and we consider, as a further example, the case $d=3$, corresponding to the M2-brane in the 11-dimensional supergravity context, for which we again obtain an additional transport coefficient not previously reported, as far as we are aware.

The explicit expressions for the transport coefficients, in the small spatial momentum $\mathfrak{q}$ regime, are found by performing an expansion of the wave solution in the bulk problem involving multiple polylogarithms in several variables \cite{Lewin'81, Goncharov:1998kja, MINH2000217, goncharov2001multiple, Wald}, that we introduce explicitly in \eqref{mpolylogs}.
The recursive structure of the method enables a systematic characterisation of the results at any order in the $\mathfrak{q}$-expansion and in any dimension $d$.

We end by discussing the relations of our results to different sectors of gravitational perturbations, and some open questions.

Using this methodological framework, we sum up the three main results we obtain: we get an explicit expression for the transport coefficients up to order $\mathfrak{q}^{10}$ in $\mathcal{N}=4$ SYM and up to order $\mathfrak{q}^{6}$ in the bulk 4-dimensional case, and we characterise  the structure of irrational numbers involved in the expression of the transport coefficients at every order $\mathfrak{q}^{2k}$ in $(d+1)$-dimensions.

Attached to the arXiv submission, the {\tt Mathematica} notebooks with the computed frequencies and wave solutions in the $d=4$ and $d=3$ cases can also be found.

\section{Shear sector in five dimensions}

The backgrounds \eqref{metricSAdS} appear as the near-horizon limit of brane backgrounds. In this section, we discuss the $d=4$ case, which can be obtained by starting from the near-horizon metric of the non-extremal three-brane background \cite{Policastro:2002se}
\begin{equation}\label{metricbrane}
\begin{aligned}
\mathrm{d}s^2_{10}=\,&\frac{1}{\sqrt{H(r)}}\left[-f(r)\,\mathrm{d}t^2+\mathrm{d}x_1^2+\mathrm{d}x_2^2+\mathrm{d}x_3^2\right]+\\
&\sqrt{H(r)}\left[\frac{\mathrm{d}r^2}{f(r)}+r^2\,\mathrm{d}\Omega^2_5\right],
\end{aligned}
\end{equation}
where
\begin{equation}
\begin{aligned}
H(r)=1+\frac{R^4}{r^4},\quad
f(r)=1-\frac{r_0^4}{r^4}.
\end{aligned}
\end{equation}
Indeed, in the limit $r\ll R$, \eqref{metricbrane} can be rewritten as
\begin{equation}\label{10dim}
\mathrm{d}s^2_{10}=\frac{(\pi T R)^2}{z}\left(-f(z)\,\mathrm{d}t^2+\mathrm{d}\vec{x}^2\right)+\frac{R^2}{4\, z^2\, f(z)}\,\mathrm{d}z^2+R^2\,\mathrm{d}\Omega^2_5,
\end{equation}
with $\mathrm{d}\vec{x}^2=\mathrm{d}x_1^2+\mathrm{d}x_2^2+\mathrm{d}x_3^2$, and
where $T=r_0/\left(\pi R^2\right)$ is the Hawking temperature, the local variable $z$ is defined as $z=r_0^2/r^2$, so that the horizon is at $z=1$ and the AdS boundary at $z=0$, and $f(z)=1-z^2$.

The background \eqref{10dim} is dual to the $\mathcal{N}=4$ $SU(N)$ SYM at finite temperature $T$, in the limit $N\to\infty$ and $N\,g_{\text{YM}}^2\to\infty$.

We consider a small off-diagonal gravitational perturbation of the five-dimensional part of the geometry \eqref{10dim}, that is we consider
$g_{\mu\nu}=g_{\mu\nu}^{(0)}+h_{\mu\nu}$, where $g_{\mu\nu}^{(0)}$ is given by
\begin{equation}
\begin{aligned}
\mathrm{d}s^2_5=\frac{(\pi T R)^2}{z}\left(-f(z)\,\mathrm{d}t^2+\mathrm{d}\vec{x}^2\right)+\frac{R^2}{4\, z^2\, f(z)}\,\mathrm{d}z^2,
\end{aligned}
\end{equation}
and where the perturbations satisfies $h_{tx_1}\ne 0$ and $h_{x_1x_3}\ne 0$. In this channel, the correlation functions admit a diffusion pole.

The Einstein equations for $h_{\mu\nu}$ provide a system of coupled differential equations for the Fourier components of $H_t=z\,h_{tx_1}/\left(\pi T R\right)^2$ and $H_{x_3}=z\,h_{x_3x_1}/\left(\pi T R\right)^2$ (see Eq. (6.13) in \cite{Policastro:2002se}).

The relevant differential equation we consider is obtained in the low-frequency and long-wavelength regime, by solving for $H_{x_3}$ and redefining the wave function $H'_t$ as
\begin{equation}
H'_t(z)=z\left(1-z^2\right)^{-\frac{i \mathfrak{w} }{2}}\,\psi(z).
\end{equation}
The resulting differential equation reads
\begin{align}
&\psi''(z)+\frac{2+2 i\, (\mathfrak{w} +2 i)\, z^2}{z-z^3}\,\psi'(z)+\label{diffeq}\\
&\frac{(1+z) \mathfrak{q}^2 -\mathfrak{w}  \left[\mathfrak{w} +\left(\mathfrak{w} +3 i\right)\left(z+1\right)z\right]}{z\left(z-1\right)\left(z+1\right)^2}\,\psi(z)=0,\nonumber
\end{align}
where $\mathfrak{w}$ and $\mathfrak{q}$ are defined in units of $2\pi T$.
This differential equation can be brought into a Heun differential equation \cite{heun1888theorie, ronveaux1995heun} with a suitable redefinition of the local variable and of the wave function, having result singular points at $z=0,\pm 1,\infty$. 
\footnote{The differential equation \eqref{diffeq} agrees with Eq. (6.16) in \cite{Policastro:2002se} by considering the redefinition of the wave function
$G(z)=z\,(1+z)^{-\frac{i \mathfrak{w} }{2}}\,\psi(z)$, and it also agrees with the differential equation in Eq. (5.9) in \cite{Kovtun:2003wp} with the parameters specified in Eq. (5.19) and $p=3$, by redefining the wave function as
$A'_t(z)=z\left(1-z^2\right)^{-i\mathfrak{w}/2}\,\psi(z)$.
}

The differential equation \eqref{diffeq} is supplied by asking regularity conditions at $z=0$ and at $z=1$, which correspond to a vanishing Dirichlet boundary condition at the AdS boundary and an ingoing boundary condition at the horizon in the original variables, respectively.

Recently, there have been important developments in the solution of the connection problems for the Heun equation and its confluence class \cite{Bonelli:2022ten, Lisovyy:2022flm}, thanks to the connection with Liouville conformal field theory and supersymmetric gauge theory. However, the position of the regular singular points in \eqref{diffeq} implies that the expansion parameter of these connection formulae, also referred to as the instanton counting parameter, assumes a finite value, and does not provide an efficient way of analytically treating the specific problem at hand.

Therefore, to compute the coefficients of the dispersion relation in the small $\mathfrak{q}$ expansion, we use a different recursive approach which generalises the one developed in \cite{Aminov:2023jve}, and that we now describe.\footnote{With respect to the Heun problem in Liouville conformal field theory, the expansion we present below corresponds to the expansion around a degenerate intermediate channel. We thank Cristoforo Iossa for pointing our attention towards this aspect.}

We start by introducing the Ansatz
\begin{align}
\mathfrak{w}&=\sum_{n\ge 1}\mathfrak{w}_n\mathfrak{q}^{2n},\label{dispersion}\\
\psi(z)&=\sum_{n\ge 0}\psi_n(z)\mathfrak{q}^{2n}.
\end{align}
and taking as the leading order solution
\begin{equation}
\psi_0(z)=1,
\end{equation}
which is regular at the horizon and at the AdS boundary. Then, we recursively construct the higher order corrections $\psi_n(z)$ and $\mathfrak{w}_n$, $n\ge 1$, starting from the basis of leading order solutions of \eqref{diffeq} in $\mathfrak{q}^2$, which is given by
\begin{equation}
\psi_0(z)=1,\quad\quad
g_0(z)=-\frac{1}{z}-\frac{1}{2} \log (1-z)+\frac{1}{2} \log (1+z).
\end{equation}
The Wronskian between these two solutions is
\begin{equation}
W_0(z)=\frac{1}{z^2-z^4}.    
\end{equation}
More precisely, let $k\in\mathbb{N}$. Suppose we have already determined $\mathfrak{w}_1,\dots,\mathfrak{w}_{k-1}$ and the corrections $\psi_1(z),\dots,\psi_{k-1}(z)$. Then, we can write the solution for $\psi_k(z)$ in the form
\begin{widetext}
\begin{equation}\label{correctionk}
\psi_k(z)=c_k\,g_0(z)-g_0(z)\int^z\frac{\eta_k(z')\psi_0(z')}{W_0(z')}\mathrm{d}z'+\psi_0(z)\int^z\frac{\eta_k(z')g_0(z')}{W_0(z')}\mathrm{d}z'.
\end{equation}
\end{widetext}
In \eqref{correctionk}, $c_k$ is an integration constant which can be fixed by requiring regularity of the solution at the horizon $z=1$, and $\eta_k(z)$ is the inhomogenous part of the differential equation at order $\mathfrak{q}^{2k}$, when the solution
\begin{equation}
\sum_{n=0}^k\psi_n(z)\mathfrak{q}^{2n}
\end{equation}
is inserted with all $\psi_0(z),\dots,\psi_{k-1}(z)$ and $\mathfrak{w}_1,\dots,\mathfrak{w}_{k-1}$ determined.
In principle, \eqref{correctionk} should depend on two integration constants, and on the right hand side one should include the additional term $b_k\,\psi_0(z)$, for some constant $b_k$. However, a nonzero $b_k$ would contribute to a different normalization of the leading order solution $\psi_0(z)$, so we have the freedom to fix these constants to 0 at each order.

At order $k=1$, the integrals in \eqref{correctionk} can be obtained easily by directly integrating. In particular, one has
\begin{equation}\label{eta1}
\eta_1(z)=\frac{1-3 i \mathfrak{w}_1 z}{z\,(z^2-1)},
\end{equation}
and, as a consequence,
\begin{align}
&\int^z\frac{\eta_1(z')\psi_0(z')}{W_0(z')}\mathrm{d}z'=\frac{1}{2} z^2 (-1+2 i \mathfrak{w}_1 z)\\
&\int^z\frac{\eta_1(z')g_0(z')}{W_0(z')}\mathrm{d}z'=\frac{1}{2} z (1-2 i \mathfrak{w}_1 z)\nonumber\\
&+\frac{1}{4} \left(-2 i \mathfrak{w}_1 \left(z^3-1\right)+z^2-1\right) \log (1-z)\nonumber\\
&+\frac{1}{4} \left(2 i \mathfrak{w}_1 \left(z^3+1\right)-z^2+1\right) \log (1+z).
\end{align}
Inserting these results in $\psi_1(z)$, we find that the solution is regular at $z=0$ only if $c_1=0$. Finally, the solution is regular at $z=1$ only if $\mathfrak{w}_1=-i/2$, fixing the first transport coefficient and the expression of the first order correction of the wave solution, which then reads $\psi_1(z)=\log (1+z)/2$.

For the higher order corrections, it is not possible to solve the integrals in \eqref{correctionk} in terms of elementary functions. Therefore, we determine the integrals, that we denote with $I_1^{(k)}(z),I_2^{(k)}(z)$, using a recursive procedure that we now describe. The idea, which will be made more precise below, is that the set of special functions that can appear at order $k$ can be deduced from the set of special functions appearing in previous orders $j\le k-1$. By building an undetermined linear combination of these special functions, we determine the integrals by asking the $z$-derivatives of these linear combinations to match the integrands in \eqref{correctionk}. By recursion, the procedure generalises to every order $k$, and only requires the knowledge of the singularity structure of the differential equation and of its leading order solutions.

We start by fixing a graded family of sets $ \mathcal{B} = (B_k)_{k \in \mathbb{Z}_{\geq 1}}$ of special functions, where each $B_k$  is a finite set of functions to which we assign degree $k$. These functions of degree $k$, together with the elements of $B_i$, with $i<k$, will be the special functions that in principle can enter the expression of $\psi_k(z)$.
The functions of degree $k$ may include products or powers of lower-degree functions (e.g., we can take $\log^2(1 - z)$ as an element of $B_2$ if $\log(1 - z)$ was an element of $B_1$), but we require that the elements of $B_k$ are linearly independent over the $\mathbb{C}(z)$-vector space
\begin{equation}
\mathrm{span}_{\mathbb{C}(z)}(B_1 \cup \cdots \cup B_{k-1}).
\end{equation}
In particular, there are different possible definitions for the set $B_k$, depending on which elements we choose to include in it. We always assume that one of these possible choices has been fixed and we provide details when needed.

We also extend the notion of degree (that for elements of $B_i$ is defined to be $i$, for every $i=1,\dots,k$) to any product of different elements as the sum of the degrees of the factors, and to a generic $\phi(z)\in\mathrm{span}_{\mathbb{C}(z)}(B_1 \cup \cdots \cup B_{k})$ as the maximum degree among the degrees of the summands appearing in $\phi(z)$.

We then build $I_1^{(k)}(z)$ and $I_2^{(k)}(z)$ as undetermined functions in $\mathrm{span}_{\mathbb{C}(z)}(B_1 \cup \cdots \cup B_{k})$ with all possible terms of degree $\le k$, with undetermined coefficients, and we then fix these coefficients by matching the $z$-derivative of $I_1^{(k)}(z)$ and $I_2^{(k)}(z)$ with the two integrands in \eqref{correctionk}.

In our concrete case, we include in $B_1$ the functions $\log(1-z)$ and $\log(1+z)$.
Then, for $k>1$, we include in our basis a class of multiple polylogarithms in several variables and products of functions of lower degree. 
We recall that the multiple polylogarithms in several variables are special functions defined by the Taylor series expansion
\begin{equation}\label{mpolylogs}
\text{Li}_{s_1,\dots,s_n}(z_1,\dots,z_n)=\sum_{k_1>k_2>\dots>k_n\ge 1}\frac{z_1^{k_1}\dots z_n^{k_n}}{k_1^{s_1}\dots k_n^{s_n}},
\end{equation}
where $n,s_1,\dots,s_n\in\mathbb{N}$. The integer $n$ is called level, and the sum $s_1+\dots+s_n$ is called weight. Moreover, they satisfy the recurrence relation
\begin{equation}
z_1 \,\partial_{z_1} \text{Li}_{s_1,\dots,s_n}(z_1, \dots, z_n)=\text{Li}_{s_1-1,\dots,s_n}(z_1, \dots, z_n),
\end{equation}
for $s_1>1$, and
\begin{equation}
(1-z_1) \,\partial_{z_1} \text{Li}_{s_1,\dots,s_n}(z_1, \dots, z_n)=\text{Li}_{s_2,\dots,s_n}(z_1 z_2,z_3, \dots, z_n),
\end{equation}
for $s_1=1$ and $n\ge 2$.

The class of multiple polylogarithms that we include in $B_k$ is a (choice of a) set of multiple polylogarithms with weight $k$ which are independent over $\mathrm{span}_{\mathbb{C}(z)}(B_1 \cup \cdots \cup B_{k-1})$, and such that the first variable $z_1$ equals $-z$ if $s_1=1$ and assume values in $\{\pm z\}$ for $s_1\ge 2$, and the other variables $z_2,\dots,z_n$ assume values in $\{\pm 1\}$. 
Finally, we also include in $B_k$ the products of functions with lower degree in $\bigcup_{j=1}^{k-1}B_j$ and with total weight $k$ which cannot be rewritten as linear combinations of the functions already included in $B_k$.  
We postpone to Appendix \ref{appMPL} the list of identities used and the elements included in our basis.

With this recursive definition, we argue that the set of functions $\mathrm{span}_{\mathbb{C}(z)}(B_1 \cup \cdots \cup B_{k})$ is enough to describe the solution $\psi_k(z)$ for every $k\ge 1$. To see this, it is useful to provide an alternative definition of the multiple polylogarithms in several variables in terms of one-forms\footnote{For $\omega_{z_1}, \dots, \omega_{z_p}$ differential one-forms, with $\omega_{z_i}=f_{z_i}(t)\mathrm{d}t$ 
for some function $f_{z_i}$, we define inductively $\int_0^x\omega_{z_1} \dots \omega_{z_p}=\int_0^x f_{z_1}(t)\mathrm{d}t \int_0^t \omega_{z_2}\dots\omega_{z_p}$.} \cite{waldschmidt2002multiple}
\begin{equation}\label{iteratedint}
\begin{aligned}
&\text{Li}_{s_1,\dots,s_n}(z_1, \dots, z_n)=\\
&\int_0^1\omega_0^{s_1-1}\omega_{z_1}\omega_0^{s_2-1}\omega_{z_1z_2}\dots\omega_0^{s_n-1}\omega_{z_1\dots z_n},
\end{aligned}
\end{equation}
where
\begin{equation}
\omega_{z}=
\left\{
\begin{aligned}
&\frac{z\mathrm{d}t}{1-z t}, & z\neq 0,\\
&\frac{\mathrm{d}t}{t}, & z= 0.
\end{aligned}
\right.
\end{equation} 
By analysing the singularity structure of the differential equation and the seed of our recursive procedure, which is specified by $\psi_0(z), g_0(z)$, and $W_0(z)$, we see that the one-forms involved in our problem are $\omega_0, \omega_1, \omega_{-1}$. Moreover, when building the results of the integrals in \eqref{correctionk}, we can consider the partial fraction decomposition of the integrands, and after every integration the total weight of the functions involved can increase at most by one, which happens for the integrals of a function in $B_{k-1}$ against one of the three 1-forms (see Appendix C of \cite{Aminov:2023jve} for details on the decomposition of the integrands).
We remark that we did not include $\log(z)$ in $B_1$, since this function never appears in our solution's expansion. The fact that $\log(z)$ does not appear in $\psi_1(z)$ is a consequence of the fact that the partial fraction decomposition of $\eta_1(z)/W_0(z)$ and $\eta_1(z)g_0(z)/W_0(z)$ do not include a term of the form $c/z$, with $c\in\mathbb{C}$, where the expression of $\eta_1(z)$ is in \eqref{eta1}.
For $k>1$, the same reasoning applies recursively, since for a term of the form $c/z$ to appear in the partial fraction decomposition of the integrals there should be such a term in the partial fraction decomposition $\eta_k(z)$, but this is avoided by imposing the regularity boundary condition at $z=0$ at order $k-1$.
Moreover, with an analogous recursive reasoning, we can actually reduce the space of the functions entering in the results $I_1^{(k)}, I_2^{(k)}$ of the integrals in \eqref{correctionk} at order $k$ to be $\mathrm{span}_{\mathbb{C}[z]}(B_1 \cup \cdots \cup B_{k})$, that is we can restrict to consider only polynomials in $z$ instead of rational functions in building the integrals of \eqref{correctionk}, with a bounded degree for the polynomials.

We stressed that a crucial point in the definition of $B_k$ is the independence of the functions entering the set. Indeed, not every functions of the form specified above are independent. Easy examples of identities are
\begin{align}
\text{Li}_{\underbrace{1,\dots,1}_{n}}(z,\underbrace{1,\dots,1}_{n-1})&=\frac{(-1)^n \log ^n(1-z)}{n!},\quad n\ge 1.
\end{align}
There are, however, more involved identities which come from integrating by parts the definition \eqref{iteratedint} or by the shuffle relations satisfied by products of multiple polylogarithms.\footnote{Multiple polylogarithms in several variables also satisfy stuffle relations, but we do not care about them because the results of the stuffle product will produce functions with variables having different prescriptions from the ones we specified.} Examples of these identities are 
\begin{equation}\label{idweight2ex}
\text{Li}_{1,1}(z,-1)=-\text{Li}_{1,1}(-z,-1)+\log (1-z) \log (1+z).
\end{equation}
and
\begin{equation}\label{shuffleex}
\text{Li}_2(z)^2=4 \text{Li}_{3,1}(z,1)+2\text{Li}_{2,2}(z,1).
\end{equation}
The identities of the form \eqref{idweight2ex} are important also to impose the boundary condition at $z=1$. Indeed, all multiple polylogarithms of the form $\text{Li}_{1,s_2,\dots,s_n}(z,z_2,\dots,z_n)$ have a logarithmic divergence at $z=1$, and we want to avoid these functions by using identities which encode all those divergences in terms of $\log(1-z)$, as in \eqref{idweight2ex}. This enables us to easily impose the boundary condition by asking the coefficient of $\log(1-z)$ in the limit $z\to 1$ in \eqref{correctionk} to be equal to zero.
This will fix the integration constant $c_k$. Finally, we determine $\mathfrak{w}_k$ by imposing the regularity of the function at $z=0$.

We here collect the results for 
the five computed results for the shear mode's transport coefficients:
\begin{align}
\mathfrak{w}_1&=-\frac{i}{2},\\
\mathfrak{w}_2&=-\frac{i}{4}(1-\log 2),\\
\mathfrak{w}_3&=-\frac{i}{96} \left[24 \log ^2 2-\pi ^2\right],
\end{align}
\begin{equation}
\begin{aligned}
\mathfrak{w}_4&=-\frac{i}{384} \Bigl\{-3 \left[7 \zeta (3)+8 (5 \log 2-3) \log ^2 2+8 \log 2\right]\\
&\ \ \ +2 \pi ^2 (5 \log 2-1)\Bigr\},
\end{aligned}
\end{equation}
\begin{equation}
\begin{aligned}
\mathfrak{w}_5&=-\frac{i}{9216}\,\Bigl\{-72 \left[8 \text{Li}_4(1/2)+3 \zeta (3)\right]+1008 \zeta (3) \log (2)\\
&\ \ \ +\frac{41 \pi ^4}{5}+4008 \log ^4(2)-4224 \log ^3 2+576 \log ^2 2\\
&\ \ \ -24 \pi ^2 \left(1+20 \log ^2 2-10 \log 2\right)\Bigr\}.
\end{aligned}
\end{equation}
The results for $\mathfrak{w}_1,\dots,\mathfrak{w}_4$ are in accordance with those in \cite{Grozdanov:2019uhi}, whereas $\mathfrak{w}_5$ is a new result.
We include the results for the wave function's solution in the ancillary file ``shear\_SAdS5.m''.
In the notebook, we denoted with \verb|integral|$k i$ the results of the integrals appearing in \eqref{correctionk} for the $k$th order solution, with $i=1,2$ for the first and second integral, respectively. Moreover, the solution $\psi_k(z)$ (with the integration constants fixed by the boundary conditions) is denoted with \verb|fz|$k(z)$. In their expressions, we denote with \verb|Li|$[\{z_1,\dots,z_n\}]$ the functions $\text{Li}_{s_1,\dots,s_n}(z_1,\dots,z_n)$ with $s_1=\dots=s_n=1$, and the multiple polylogarithms in several variables with some $s_i\ne 0$ with \verb|Li|$[\{s_1,\dots,s_n\},\{z_1,\dots,z_n\}]$.

In general, because of the structure of the solution at order $k$, we have control on the irrational numbers that can enter the expression of $\mathfrak{w}_k$. Indeed, as said before, these can be found by imposing the regularity of the wave function at $z=0$. Since all multiple polylogarithms of weight $k$ enter with constant coefficients in $\psi_k(z)$, these vanish at $z=0$ because of their Taylor series expansion. The only irrational numbers come from the evaluation of the remaining part of the wave function at $z=0$ and from the integration constant $c_k$, which was found by imposing the coefficient of $\log(1-z)$ in the limit $z\to 1$ to equal zero. This implies that $c_k$, and so also $\mathfrak{w}_k$, admits an expression in terms of product of multiple polylogarithms evaluated at $z=1$. We conclude that the irrational numbers that can appear are $\log 2$, $\zeta$-values of order at most $k-1$, or, in general, colored multiple zeta values of level 2 and weight $\le k-1$ \cite{xu2024apery}.

\section{The general \texorpdfstring{$d$}{} construction}

The analysis presented in the previous section for $d=4$ can be extended to the shear sector in different dimensions.

More precisely, the singularity structure of the same problem for planar SAdS$_{d+1}$ is determined by the singularities of the blackening factor
\begin{equation}
f(r)=r^2-\frac{1}{r^{d-2}},
\end{equation}
where we took the normalisation $\mu=R=1$, so that the horizon is at $r=1$. 

If $d$ is odd, then $r^{d-2}\,f(r)=0$ has $d$ roots, which correspond to the $d$th roots of unity. Including the AdS boundary $r=\infty$ and the origin $r=0$, the singularities of the spectral problem are $d+2$ regular singular points. If $d$ is even, instead, it is possible to redefine the variable as $z=1/r^2$ as in the previous section, and $z^{\frac{d-2}{2}}f(z)=0$ has $d/2$ roots, which correspond to the $(d/2)$th roots of unity. Together with the AdS boundary at $z=0$ and the origin $z=\infty$, the regular singularities of the spectral problem are $d/2+2$.

Generalising the case seen in the previous section, the same spectral problem appears in the $\mathcal{N}=2$ $SU(2)^{d-1}$ (for odd $d$) or $SU(2)^{d/2-1}$ (for even $d$) superconformal linear quiver gauge theory. The connection problems for these class of theories were analysed in \cite{Arnaudo:2025kof}.
However, as for the Heun case for $d=4$, also in these cases the instanton counting parameters assume finite values, and do not provide an efficient expansion for the frequency.

Instead, the procedure described with an expansion in terms of multiple polylogarithms in several variables can be straightforwardly generalised, by adapting the construction of the basis $\mathcal{B}$ to the presence of additional singular points.

More explicitly, we include in $B_1$ the functions
\begin{equation}
\log(1-u_i\,z),\quad i=\begin{cases}
1,\dots d,\ \text{if}\ d\ \text{odd},\\
1,\dots \frac{d}{2},\ \text{if}\ d\ \text{even},
\end{cases}
\end{equation}
where $u_i$ are the roots of unity (including $u_d=1$ or $u_{d/2}=1$, for odd or even $d$, respectively).

Then, for $k>1$, we include in the basis the class of multiple polylogarithms in several variables with weight $k$ which are independent over $\mathrm{span}_{\mathbb{C}[z]}\left(B_1\cup\dots\cup B_{k-1}\right)$ such that the first variable assume values in $\{u_i z\}$ with $i=1,\dots,d-1$ (or $i=1,\dots,d/2-1$) if $s_1=1$ or in $\{u_i z\}$ with $i=1,\dots,d$ (or $i=1,\dots,d/2$) for $d$ odd (or $d$ even), respectively, and the other variables assume values in $\{u_i\}$ with $i=1,\dots,d$ (or $i=1,\dots,d/2$) for $d$ odd (or $d$ even), respectively.
In addition to these functions, as in the $d=4$ case, we also add the products of functions of lower degree in $\bigcup_{j=1}^{k-1}B_j$ with total weight $k$ which cannot be rewritten as linear combinations of the functions already included in $B_k$.

The generalisation of the results obtained in the $d=4$ case implies that the transport coefficients $\mathfrak{w}_k$ for the $(d+1)$-dimensional brane admit an expression in terms of colored multiple zeta values of weight $\le k-1$ and level $d$ for odd $d$ and level $d/2$ for even $d$, and the set of irrational numbers involved in the expressions of $\mathfrak{w}_k$ is a subset of those appearing in the corresponding set of colored multiple zeta values.

This discussion also explains why the $d=4$ case is the simplest one, since it involves the fewer number of special functions to be introduced.

\subsection{Shear sector in four dimensions}

As a further example, we report the results for the case of SAdS$_4$, where, as just discussed, we expect the transport coefficients to be expressed in terms of multiple polylogarithms evaluated at third roots of unity.

The shear mode perturbation equation for SAdS$_4$ can be read from Eq. (5.9) in \cite{Kovtun:2003wp} by taking the parameters in Eq. (5.19) and $p=1$:
\begin{equation}
\frac{\mathrm{d}}{\mathrm{d}z}\left[\left(1-z^3\right)z^2\frac{\mathrm{d}}{\mathrm{d}z}\left(z^{-2}A'_t(z)\right)\right]+\left(\frac{\mathfrak{w}^2}{1-z^3}-\mathfrak{q}^2\right)A'_t(z)=0.
\end{equation}
By redefining the wave function as
\begin{equation}
A'_t(z)=z^2 \left(1-z^3\right)^{-\frac{i \mathfrak{w}}{2}}\psi(z),
\end{equation}
the relevant differential equation reads 
\begin{equation}
\begin{aligned}
&\psi''(z)+\frac{(5-3 i \mathfrak{w} ) z^3-2}{z \left(z^3-1\right)}\psi'(z)+\\
&\left[\frac{9 \mathfrak{w} ^2}{(4-4 z) \left(z^2+z+1\right)^2}+\frac{9 \mathfrak{q}^2-3 \mathfrak{w}  (3 \mathfrak{w} +8 i) z}{4 (z-1) \left(z^2+z+1\right)}\right]\psi(z)=0.
\end{aligned}
\end{equation}
A basis of leading order solutions is given by 
\begin{equation}
\begin{aligned}
&\psi_0(z)=1,\\
&g_0(z)=-u_1 \log \left(1-u_1 z\right)-u_2 \log \left(1-u_2 z\right)\\
&\quad\quad\quad\ -\frac{3}{z}-\log (1-z),
\end{aligned}
\end{equation}
and the Wronskian between the two solutions is 
\begin{equation}
W_0(z)=\frac{3}{z^2-z^5}.
\end{equation}
Using the same Ansatz \eqref{dispersion},
we computed the results up to order $\mathfrak{q}^6$.
Using the notation
\begin{equation}
u_1=-\frac{1}{2}-i\,\frac{\sqrt{3}}{2},\qquad u_2=-\frac{1}{2}+i\,\frac{\sqrt{3}}{2},
\end{equation}
for the two nontrivial third roots of unity, we include in the basis the functions
\begin{equation}
\text{Li}_{1,1}\left(u_1z,u_2\right),\quad \text{Li}_{1,1}\left(u_1z,u_1\right),\quad \text{Li}_{1,1}\left(u_2z,u_1\right)
\end{equation}
in $B_2$, and 
\begin{equation}
\begin{aligned}
&\text{Li}_{1,1,1}\left(u_1\,z,1,u_2\right),\quad \text{Li}_{1,1,1}\left(u_1\,z,1,u_1\right),\quad \text{Li}_{1,1,1}\left(u_1\,z,u_2,1\right),\\ &\text{Li}_{1,1,1}\left(u_1\,z,u_1,u_1\right),\quad
\text{Li}_{1,1,1}\left(u_1\,z,u_1,1\right),\quad \text{Li}_{1,1,1}\left(u_2\,z,1,u_1\right),\\ &\text{Li}_{1,1,1}\left(u_2\,z,u_1,1\right),\quad \text{Li}_{1,1,1}\left(u_2\,z,u_2,u_2\right)
\end{aligned}
\end{equation}
in $B_3$. The relevant identities among this set of functions were analysed in Appendix D in \cite{Aminov:2023jve}.

We include the results for the first-order wave function
\begin{widetext}
\begin{equation}
\begin{aligned}
\psi_1(z)=&\,\frac{1}{48} \biggl[\frac{12 \left(z-\log \left(1-u_1 z\right)\right)}{u_1^2}+\frac{12 \log \left(1-u_1 z\right)}{u_1^3}+\frac{12 \left(z-\log \left(1-u_2 z\right)\right)}{u_2^2}+\frac{12 \log \left(1-u_2 z\right)}{u_2^3}\\
&+u_1 (3 z-4) z^3+u_2 (3 z-4) z^3+\frac{6 (z-2) z}{u_1}+\frac{6 (z-2) z}{u_2}+\left(3 z^2-4 z+6\right) z^2\biggr],
\end{aligned}
\end{equation}
\end{widetext}
and for the first three transport coefficients
\begin{equation}
\begin{aligned}
\mathfrak{w}_1&=-\frac{i}{2},\\
\mathfrak{w}_2&=-\frac{i}{48} \left(\pi  \sqrt{3}+9-9 \log (3)\right),
\end{aligned}
\end{equation}
\begin{widetext}
\begin{equation}
\begin{aligned}
\mathfrak{w}_3=&\,\frac{1}{1152}\Bigl\{72 \sqrt{3} \text{Li}_{1,1}\left(u_1,u_1\right)+36 \sqrt{3} \left(u_1-u_2-1\right) \text{Li}_{1,1}\left(u_2,u_1\right)+36 \sqrt{3} \left(u_1-u_2+1\right) \text{Li}_{1,1}\left(u_1,u_2\right)\\
&+\left(u_1-u_2\right) \left(\pi ^2 \sqrt{3}-9 i \log ^2(3)-6 \pi  (3+4 \log (3))\right)-\sqrt{3}\,\pi ^2-27 i (12+\log (3) (5 \log (3)-6))\Bigr\}.
\end{aligned}
\end{equation}
\end{widetext}
The higher order results for the wave solutions are included in the ancillary file ``shear\_SAdS4.m'', where the notations agree with the ones discussed in the $d=4$ case.

We finally remark that the first two computed coefficients are in agreement with the results in Table III in \cite{Natsuume:2007ty}, whereas $\mathfrak{w}_3$ is a new result.

\section{Discussion}

The above analysis was presented for the shear sector of gravitational perturbations. In principle, the same line of reasoning could be extended to the other sectors, like the sound one. However, in this case an additional complication arises: it is not possible to impose both boundary conditions -- at the horizon and at the AdS boundary -- on a single local solution. This obstruction originates from the more intricate singularity structure of the corresponding differential equation. 
More precisely, starting from the notations of \cite{Kodama:2003jz} for the scalar sector of gravitational perturbations, if there are singular points scaling with different powers of the black brane mass parameter in the original $r$ variable, local analyses (in the near-horizon region or in the near-boundary region) will make some of the singularities to overlap, and additional regions must be introduced. 
This can be seen explicitly, since a solution expanded around one of the two points where the boundary conditions are prescribed does not converge in the vicinity of the other.
The implementation of this procedure in the four-dimensional case has been examined in detail in \cite{Aminov:2023jve}.\footnote{Nonetheless, we expect that the conclusion regarding the possible irrational numbers appearing in the expressions of the QNMs should hold for the sound sector as well.}
Moreover, the same procedure we outlined can be applied to perturbations of D$p$-branes, as the differential equations describing them share the same structure \cite{Kovtun:2004de, Natsuume:2007ty}.

Even in the context of the shear sector, it would be interesting to make contact with results in the large $d$ expansion \cite{Emparan:2013moa, Emparan:2013xia, Emparan:2015rva, Emparan:2015hwa, Andrade:2018zeb, Grozdanov:2023tag}. Indeed, in this limit the gravitational dynamics exhibit a significant simplification: the perturbations become strongly localized near the horizon, and the quasinormal frequencies admit a leading diffusive behavior with systematically computable corrections. From the point of view of the presented method, increasing $d$ leads to the presence of more singularities, which in the $z$ variable and in the $d\to\infty$ limit will densely fill the unit circumference.
Since the simplification of the $1/d$ expansion is not manifest in our description, its precise emergence remains to be understood.

Finally, we believe the techniques presented in this work related to expansion in terms of multiple polylogarthms apply to more general contexts.
Indeed, multiple polylogarithms have emerged as a central class of special functions in a variety of physical contexts, ranging from multiloop Feynman integrals in quantum field theory \cite{Henn:2013pwa, Panzer:2015ida}, to perturbative gravitational and string amplitudes in AdS \cite{Alday:2024yax, Alday:2024ksp, Alday:2024rjs}. 
This highlights the utility of multiple polylogarithms as a powerful algebraic framework for organizing analytic structures, with their intrinsic algebraic features enabling the systematic representation and manipulation of nontrivial functional relations.
\\

\vspace{0.5cm}

\begin{acknowledgments}
It is a pleasure to thank Gleb Aminov and Benjamin Withers for useful discussions. The author is supported by the Royal Society grant URF{\textbackslash}R{\textbackslash}231002, ‘Dynamics of holographic field theories’.
\end{acknowledgments}

\appendix

\begin{widetext}

\section{Multiple polylogarithms basis and identities for \texorpdfstring{$d=4$}{}}\label{appMPL}

In this Appendix, we list the identities used for selecting the set of special functions to include in our basis. 

We begin by discussing the case of multiple polylogarithms in several variables of the form
\begin{equation}
\text{Li}_{\underbrace{1,\dots,1}_{n}}(\theta_1\,z,\theta_2,\dots,\theta_n),\quad \theta_i\in\{-1,1\}\quad\forall\,i=1,\dots,n
\end{equation}
up to weight $n=5$.

We chose to include the following functions in our basis: \begin{equation}
\begin{aligned}
\text{Li}_{1,1}(-z,-1)
\end{aligned}
\end{equation}
at weight 2, 
\begin{equation}
\begin{aligned}\text{Li}_{1,1,1}(-z,1,-1), \text{Li}_{1,1,1}(-z,-1,1)
\end{aligned}
\end{equation}
at weight 3, \begin{equation}
\begin{aligned}
\text{Li}_{1,1,1,1}(-z,-1,1,1), \text{Li}_{1,1,1,1}(-z,1,-1,1), \text{Li}_{1,1,1,1}(-z,-1,1,-1), \text{Li}_{1,1,1,1}(-z,1,1,-1)
\end{aligned}
\end{equation}
at weight 4, and
\begin{equation}
\begin{aligned}
& \text{Li}_{1,1,1,1,1}(-z,1,-1,1,1),\text{Li}_{1,1,1,1,1}(-z,1,-1,-1,1),\text{Li}_{1,1,1,1,1}(-z,-1,-1,1,-1),\text{Li}_{1,1,1,1,1}(-z,-1,1,-1,-1),\\
&\text{Li}_{1,1,1,1,1}(-z,-1,1,1,-1),\text{Li}_{1,1,1,1,1}(-z,-1,1,1,1),\text{Li}_{1,1,1,1,1}(-z,1,1,-1,1),\text{Li}_{1,1,1,1,1}(-z,-1,1,-1,1).
\end{aligned}
\end{equation}
at weight 5.
At level $k$, we include $2^{k-1}$ functions of this form in the basis.

We then list the useful identities up to weight 5.

Weight 2:
\begin{equation*}
\begin{aligned}
\text{Li}_{1,1}(z,-1)&=\log (1-z) \log (1+z)-\text{Li}_{1,1}(-z,-1),\\
\text{Li}_{1,1}(z,1)&=\frac{1}{2} \log ^2(1-z),\\
\text{Li}_{1,1}(-z,1)&=\frac{1}{2} \log ^2(1+z).
\end{aligned}
\end{equation*}
Weight 3:
\begin{equation*}
\begin{aligned}
\text{Li}_{1,1,1}(z,1,1)&=-\frac{1}{6} \log ^3(1-z),\\
\text{Li}_{1,1,1}(z,1,-1)&=\text{Li}_{1,1,1}(-z,-1,1)+\log (1-z) \text{Li}_{1,1}(-z,-1)-\frac{1}{2} \log (1+z) \log ^2(1-z),\\
\text{Li}_{1,1,1}(z,-1,1)&=\text{Li}_{1,1,1}(-z,1,-1)+\log (1+z) \text{Li}_{1,1}(-z,-1)-\frac{1}{2} \log (1-z) \log ^2(1+z),\\
\text{Li}_{1,1,1}(z,-1,-1)&=-2 \text{Li}_{1,1,1}(-z,-1,1)-\log (1-z) \text{Li}_{1,1}(-z,-1),\\
\text{Li}_{1,1,1}(-z,1,1)&=-\frac{1}{6} \log ^3(1+z),\\
\text{Li}_{1,1,1}(-z,-1,-1)&=-2 \text{Li}_{1,1,1}(-z,1,-1)-\log (1+z) \text{Li}_{1,1}(-z,-1).
\end{aligned}
\end{equation*}
Weight 4:
\begin{equation*}
\begin{aligned}
\text{Li}_{1,1,1,1}(z,1,1,1)&=\frac{1}{24} \log ^4(1-z),\\
\text{Li}_{1,1,1,1}(-z,1,1,1)&=\frac{1}{24} \log ^4(1+z),\\
\text{Li}_{1,1,1,1}(z,1,1,-1)&=-\text{Li}_{1,1,1,1}(-z,-1,1,1)-\frac{1}{2} \log ^2(1-z) \text{Li}_{1,1}(-z,-1)-\log (1-z) \text{Li}_{1,1,1}(-z,-1,1)\\
&\ \ \ +\frac{1}{6} \log (1+z) \log ^3(1-z),\\
\text{Li}_{1,1,1,1}(z,1,-1,-1)&=3 \text{Li}_{1,1,1,1}(-z,-1,1,1)+\frac{1}{2} \log ^2(1-z) \text{Li}_{1,1}(-z,-1)+2 \log (1-z) \text{Li}_{1,1,1}(-z,-1,1),\\
\text{Li}_{1,1,1,1}(z,-1,-1,1)&=-3 \text{Li}_{1,1,1,1}(-z,-1,1,1)-\log (1-z) \text{Li}_{1,1,1}(-z,-1,1),\\
\text{Li}_{1,1,1,1}(z,-1,-1,-1)&=-\text{Li}_{1,1,1,1}(-z,-1,1,-1)+2 \text{Li}_{1,1,1,1}(-z,1,-1,1)+\log (1+z) \text{Li}_{1,1,1}(-z,-1,1)\\
&\ \ \ +\log (1-z) \left[2 \text{Li}_{1,1,1}(-z,1,-1)+\log (1+z) \text{Li}_{1,1}(-z,-1)\right],\\
\text{Li}_{1,1,1,1}(z,-1,1,-1)&=\text{Li}_{1,1,1,1}(-z,-1,1,-1)-\log (1-z) \text{Li}_{1,1,1}(-z,1,-1)+\log (1+z) \text{Li}_{1,1,1}(-z,-1,1),\\
\text{Li}_{1,1,1,1}(z,1,-1,1)&=\frac{1}{4} \log (1+z) \left(-4 \text{Li}_{1,1,1}(-z,-1,1)-4 \log (1-z) \text{Li}_{1,1}(-z,-1)+\log (1+z) \log ^2(1-z)\right)\\
&\ \ -\text{Li}_{1,1,1,1}(-z,1,-1,1)-\log (1-z) \text{Li}_{1,1,1}(-z,1,-1),\\
\text{Li}_{1,1,1,1}(-z,-1,-1,-1)&=-\text{Li}_{1,1,1,1}(-z,-1,1,-1)-2 \text{Li}_{1,1,1,1}(-z,1,-1,1)-\log (1+z) \text{Li}_{1,1,1}(-z,-1,1),\\
\text{Li}_{1,1,1,1}(z,-1,1,1)&=\frac{1}{6} \log (1+z) (\log (1+z) (\log (1-z) \log (1+z)-3 \text{Li}_{1,1}(-z,-1))-6 \text{Li}_{1,1,1}(-z,1,-1))\\
&\ \ \ -\text{Li}_{1,1,1,1}(-z,1,1,-1),\\
\text{Li}_{1,1,1,1}(-z,-1,-1,1)&=3 \text{Li}_{1,1,1,1}(-z,1,1,-1)+\frac{1}{2} \log (1+z) (4 \text{Li}_{1,1,1}(-z,1,-1)+\log (1+z) \text{Li}_{1,1}(-z,-1)),\\
\text{Li}_{1,1,1,1}(-z,1,-1,-1)&=-3 \text{Li}_{1,1,1,1}(-z,1,1,-1)-\log (1+z) \text{Li}_{1,1}(-z,1,-1).
\end{aligned}
\end{equation*}
Weight 5: 
\begin{equation*}
\begin{aligned}
\text{Li}_{1,1,1,1,1}(-z,1,-1,1,-1)&= \frac{1}{4} \biggl[2 \text{Li}_{1,1,1,1,1}(-z,-1,-1,1,-1)-2 \text{Li}_{1,1,1,1,1}(-z,-1,1,-1,1)-6 \text{Li}_{1,1,1,1,1}(-z,1,1,-1,1)\\
&\ \ \ -\log ^2(1+z)\text{Li}_{1,1,1}(-z,-1,1)-2 \log(1+z) \text{Li}_{1,1,1,1}(-z,-1,1,-1)\\
&\ \ \ -4 \log(1+z) \text{Li}_{1,1,1,1}(-z,1,-1,1)\biggr],
\end{aligned}
\end{equation*}
\begin{equation*}
\begin{aligned}
\text{Li}_{1,1,1,1,1}(z,-1,-1,1,-1)&= -\text{Li}_{1,1,1,1,1}(-z,-1,1,-1,-1)-3 \text{Li}_{1,1,1,1,1}(-z,-1,1,1,-1)-\log (1-z) \text{Li}_{1,1,1,1}(-z,-1,1,-1),
\end{aligned}
\end{equation*}
\begin{equation*}
\begin{aligned}
\text{Li}_{1,1,1,1,1}(z,1,1,-1,1)&= \text{Li}_{1,1,1,1,1}(-z,1,-1,1,1)+\frac{1}{2} \log(1+z) \log ^2(1-z) \text{Li}_{1,1}(-z,-1)\\
&\ \ \ +\frac{1}{2} \log ^2(1-z) \text{Li}_{1,1,1}(-z,1,-1)+\log(1+z) \log (1-z) \text{Li}_{1,1,1}(-z,-1,1)\\
&\ \ \ +\log (1-z) \text{Li}_{1,1,1,1}(-z,1,-1,1)+\log(1+z) \text{Li}_{1,1,1,1}(-z,-1,1,1)-\frac{1}{12} \log ^2(1+z) \log ^3(1-z),
\end{aligned}
\end{equation*}
\begin{equation*}
\begin{aligned}
\text{Li}_{1,1,1,1,1}(z,1,-1,1,-1)&= -\text{Li}_{1,1,1,1,1}(-z,-1,1,-1,-1)-2 \text{Li}_{1,1,1,1,1}(-z,-1,1,1,-1)-\text{Li}_{1,1,1,1,1}(-z,1,-1,1,1)\\
&\ \ \ +\frac{1}{2} \log ^2(1-z) \text{Li}_{1,1,1}(-z,1,-1)-\log (1-z) \text{Li}_{1,1,1,1}(-z,-1,1,-1)\\
&\ \ \ -\log(1+z) \log (1-z) \text{Li}_{1,1,1}(-z,-1,1)-2 \log(1+z) \text{Li}_{1,1,1,1}(-z,-1,1,1),
\end{aligned}
\end{equation*}
\begin{equation*}
\begin{aligned}
\text{Li}_{1,1,1,1,1}(-z,-1,-1,-1,1)&= -\text{Li}_{1,1,1,1,1}(-z,-1,1,-1,-1)-\text{Li}_{1,1,1,1,1}(-z,-1,1,1,-1)-2 \text{Li}_{1,1,1,1,1}(-z,1,-1,1,1)\\
&\ \ \ -\log(1+z) \text{Li}_{1,1,1,1}(-z,-1,1,1),
\end{aligned}
\end{equation*}
\begin{equation*}
\begin{aligned}
\text{Li}_{1,1,1,1,1}(z,-1,1,-1,1)&= \text{Li}_{1,1,1,1,1}(-z,-1,1,-1,-1)+\text{Li}_{1,1,1,1,1}(-z,-1,1,1,-1)-\text{Li}_{1,1,1,1,1}(-z,1,-1,1,1)\\
&\ \ \ -\log (1-z) \text{Li}_{1,1,1,1}(-z,1,-1,1)+\log(1+z) \text{Li}_{1,1,1,1}(-z,-1,1,1),
\end{aligned}
\end{equation*}
\begin{equation*}
\begin{aligned}
\text{Li}_{1,1,1,1,1}(z,1,-1,-1,-1)&= \text{Li}_{1,1,1,1,1}(-z,-1,1,-1,-1)+2 \text{Li}_{1,1,1,1,1}(-z,-1,1,1,-1)-2 \text{Li}_{1,1,1,1,1}(-z,1,-1,1,1)\\
&\ \ \ -\log ^2(1-z) \text{Li}_{1,1,1}(-z,1,-1)-\frac{1}{2} \log(1+z) \log ^2(1-z) \text{Li}_{1,1}(-z,-1)\\
&\ \ \ +\log (1-z) \text{Li}_{1,1,1,1}(-z,-1,1,-1)-2 \log (1-z) \text{Li}_{1,1,1,1}(-z,1,-1,1)\\
&\ \ \ -\log(1+z) \log (1-z) \text{Li}_{1,1,1}(-z,-1,1)-\log(1+z) \text{Li}_{1,1,1,1}(-z,-1,1,1),
\end{aligned}
\end{equation*}
\begin{equation*}
\begin{aligned}
\text{Li}_{1,1,1,1,1}(z,1,-1,1,1)&= \text{Li}_{1,1,1,1,1}(-z,1,1,-1,1)+\log (1-z) \text{Li}_{1,1,1,1}(-z,1,1,-1)\\
&\ \ \ +\frac{1}{12} \log(1+z) \biggl[12 \text{Li}_{1,1,1,1}(-z,1,-1,1)+\log(1+z) \biggl(6 \text{Li}_{1,1,1}(-z,-1,1)\\
&\ \ \ +6 \log (1-z) \text{Li}_{1,1}(-z,-1)-\log(1+z) \log ^2(1-z)\biggr)+12 \log (1-z) \text{Li}_{1,1,1}(-z,1,-1)\biggr],
\end{aligned}
\end{equation*}
\begin{equation*}
\begin{aligned}
\text{Li}_{1,1,1,1,1}(z,-1,1,1,-1)&= \frac{1}{4} \biggl[-2 \text{Li}_{1,1,1,1,1}(-z,-1,-1,1,-1)-2 \text{Li}_{1,1,1,1,1}(-z,-1,1,-1,1)-\log ^2(1+z) \text{Li}_{1,1,1}(-z,-1,1)\\
&\ \ \ -2 \text{Li}_{1,1,1,1,1}(-z,1,1,-1,1)-2 \log(1+z) \text{Li}_{1,1,1,1}(-z,-1,1,-1)-4 \log (1-z) \text{Li}_{1,1,1,1}(-z,1,1,-1)\biggr],
\end{aligned}
\end{equation*}
\begin{equation*}
\begin{aligned}
\text{Li}_{1,1,1,1,1}(z,-1,1,-1,-1)&= \frac{1}{4} \biggl[2 \text{Li}_{1,1,1,1,1}(-z,-1,-1,1,-1)+6 \text{Li}_{1,1,1,1,1}(-z,-1,1,-1,1)+6 \text{Li}_{1,1,1,1,1}(-z,1,1,-1,1)\\
&\ \ \ -\log ^2(1+z) \text{Li}_{1,1,1}(-z,-1,1)+2 \log(1+z) \text{Li}_{1,1,1,1}(-z,-1,1,-1)\\
&\ \ \ +4 \log (1-z) \log(1+z) \text{Li}_{1,1,1}(-z,1,-1)+12 \log (1-z) \text{Li}_{1,1,1,1}(-z,1,1,-1)\biggr],
\end{aligned}
\end{equation*}
\begin{equation*}
\begin{aligned}
\text{Li}_{1,1,1,1,1}(-z,1,-1,-1,-1)&= \frac{1}{4} \biggl[-2 \text{Li}_{1,1,1,1,1}(-z,-1,-1,1,-1)+2 \text{Li}_{1,1,1,1,1}(-z,-1,1,-1,1)-6 \text{Li}_{1,1,1,1,1}(-z,1,1,-1,1)\\
&\ \ \ +\log ^2(1+z) \text{Li}_{1,1,1}(-z,-1,1)+2 \log(1+z) \text{Li}_{1,1,1,1}(-z,-1,1,-1)\biggr],
\end{aligned}
\end{equation*}
\begin{equation*}
\begin{aligned}
\text{Li}_{1,1,1,1,1}(z,-1,-1,-1,1)&= \frac{1}{2} \biggl[-2 \text{Li}_{1,1,1,1,1}(-z,-1,1,-1,1)-6 \text{Li}_{1,1,1,1,1}(-z,1,1,-1,1)-\log ^2(1+z) \text{Li}_{1,1,1}(-z,-1,1)\\
&\ \ \ -\log (1-z) \log ^2(1+z) \text{Li}_{1,1}(-z,-1)-4 \log(1+z) \text{Li}_{1,1,1,1}(-z,1,-1,1)\\
&\ \ \ -4 \log (1-z) \log(1+z) \text{Li}_{1,1,1}(-z,1,-1)-6 \log (1-z) \text{Li}_{1,1,1,1}(-z,1,1,-1)\biggr],
\end{aligned}
\end{equation*}
\begin{equation*}
\begin{aligned}
\text{Li}_{1,1,1,1,1}(z,-1,1,1,1)&= \frac{1}{24} \biggl[4 \text{Li}_{1,1,1,1,1}(-z,1,-1,-1,1)+\log(1+z) \biggl(12 \text{Li}_{1,1,1,1}(-z,1,1,-1)\\
&\ \ \ +\log(1+z) \left(10 \text{Li}_{1,1,1}(-z,1,-1)+4 \log(1+z) \text{Li}_{1,1}(-z,-1)-\log (1-z) \log ^2(1+z)\right)\biggr)\biggr],
\end{aligned}
\end{equation*}
\begin{equation*}
\begin{aligned}
\text{Li}_{1,1,1,1,1}(z,1,1,1,-1)&= \frac{1}{24} \biggl[4 \text{Li}_{1,1,1,1,1}(z,1,-1,-1,1)+\log (1-z)\biggl(12 \text{Li}_{1,1,1,1}(-z,-1,1,1)\\
&\ \ \ +\log (1-z) \left(10 \text{Li}_{1,1,1}(-z,-1,1)+4 \log (1-z) \text{Li}_{1,1}(-z,-1)-\log(1+z) \log ^2(1-z)\right)\biggr)\biggr],
\end{aligned}
\end{equation*}
\begin{equation*}
\begin{aligned}
\text{Li}_{1,1,1,1,1}(-z,1,1,-1,-1)&= \frac{1}{3} (\log(1+z) (3 \text{Li}_{1,1,1,1}(-z,1,1,-1)+\log(1+z) \text{Li}_{1,1,1}(-z,1,-1))-2 \text{Li}_{1,1,1,1,1}(-z,1,-1,-1,1)),
\end{aligned}
\end{equation*}
\begin{equation*}
\begin{aligned}
\text{Li}_{1,1,1,1,1}(-z,-1,-1,1,1)&= \frac{1}{6} \biggl[-4 \text{Li}_{1,1,1,1,1}(-z,1,-1,-1,1)-\log(1+z) \biggl(6 \text{Li}_{1,1,1,1}(-z,1,1,-1)\\
&\ \ \ +\log(1+z) \left(4 \text{Li}_{1,1,1}(-z,1,-1)+\log(1+z) \text{Li}_{1,1}(-z,-1)\right)\biggr)\biggr],
\end{aligned}
\end{equation*}
\begin{equation*}
\begin{aligned}
\text{Li}_{1,1,1,1,1}(-z,1,1,1,-1)&= \frac{1}{12} \bigl[2 \text{Li}_{1,1,1,1,1}(-z,1,-1,-1,1)-\log(1+z) \bigl(6 \text{Li}_{1,1,1,1}(-z,1,1,-1)+\log(1+z) \text{Li}_{1,1,1}(-z,1,-1)\bigr)\bigr],
\end{aligned}
\end{equation*}
\begin{equation*}
\begin{aligned}
\text{Li}_{1,1,1,1,1}(z,-1,-1,-1,-1)&= 2 \text{Li}_{1,1,1,1,1}(-z,-1,1,1,-1)+4 \text{Li}_{1,1,1,1,1}(-z,1,-1,1,1)+\log (1-z) \text{Li}_{1,1,1,1}(-z,-1,1,-1)\\
&\ \ \ +2 \log (1-z) \text{Li}_{1,1,1,1}(-z,1,-1,1)+2 \log(1+z) \text{Li}_{1,1,1,1}(-z,-1,1,1)\\
&\ \ \ +\log (1-z) \log(1+z) \text{Li}_{1,1,1}(-z,-1,1),
\end{aligned}
\end{equation*}
\begin{equation*}
\begin{aligned}
\text{Li}_{1,1,1,1,1}(-z,-1,-1,-1,-1)&= -\text{Li}_{1,1,1,1,1}(-z,-1,-1,1,-1)-\text{Li}_{1,1,1,1,1}(-z,-1,1,-1,1)+3 \text{Li}_{1,1,1,1,1}(-z,1,1,-1,1)\\ 
&\ \ \ +\frac{1}{2} \log ^2(1+z) \text{Li}_{1,1,1}(-z,-1,1)+2 \log(1+z) \text{Li}_{1,1,1,1}(-z,1,-1,1),
\end{aligned}
\end{equation*}
\begin{equation*}
\begin{aligned}
\text{Li}_{1,1,1,1,1}(z,1,-1,-1,1)&= 6 \text{Li}_{1,1,1,1,1}(-z,-1,1,1,1)+\frac{1}{2} \log (1-z) (6 \text{Li}_{1,1,1,1}(-z,-1,1,1)+\log (1-z) \text{Li}_{1,1,1}(-z,-1,1)),
\end{aligned}
\end{equation*}
\begin{equation*}
\begin{aligned}
\text{Li}_{1,1,1,1,1}(z,1,1,-1,-1)&= -4 \text{Li}_{1,1,1,1,1}(-z,-1,1,1,1)-\frac{1}{6} \log (1-z) \biggl[18 \text{Li}_{1,1,1,1}(-z,-1,1,1)\\
&\ \ \ +\log (1-z) \bigl(6 \text{Li}_{1,1,1}(-z,-1,1)+\log (1-z) \text{Li}_{1,1}(-z,-1)\bigr)\biggr],
\end{aligned}
\end{equation*}
\begin{equation*}
\begin{aligned}
\text{Li}_{1,1,1,1,1}(z,-1,-1,1,1)&= -4 \text{Li}_{1,1,1,1,1}(-z,-1,1,1,1)-\log (1-z) \text{Li}_{1,1,1,1}(-z,-1,1,1).
\end{aligned}
\end{equation*}

We then list the used identities that relate multiple polylogarithms having some of the $s_i$ different from 1. The functions included in our basis will then be the ones in the right hand sides of the identities.

At weight 3:
\begin{equation}
\begin{aligned}
\text{Li}_{1,2}(z,1)&=-2 \text{Li}_{2,1}(z,1)-\text{Li}_2(z) \log(1-z),\\
\text{Li}_{1,2}(-z,1)&=-2 \text{Li}_{2,1}(-z,1)-\text{Li}_2(-z) \log (1+z),\\
\text{Li}_{1,2}(z,-1)&=-\text{Li}_{2,1}(-z,-1)-\text{Li}_{2,1}(z,-1)-\text{Li}_2(-z) \log(1-z),\\
\text{Li}_{1,2}(-z,-1)&=-\text{Li}_{2,1}(-z,-1)-\text{Li}_{2,1}(z,-1)-\text{Li}_2(z) \log(1+z),
\end{aligned}
\end{equation}
and at weight 4:
\begin{equation*}
\begin{aligned}
\text{Li}_{1,2,1}(z,1,1)&=-3 \text{Li}_{2,1,1}(z,1,1)-\text{Li}_{2,1}(z,1) \log (1-z),\\ 
\text{Li}_{1,1,2}(z,1,1)&=\frac{1}{2} \left( \text{Li}_2(z) \log ^2(1-z)+4 \text{Li}_{2,1}(z,1) \log (1-z)+6 \text{Li}_{2,1,1}(z,1,1)\right),\\ \text{Li}_{1,2,1}(-z,1,1)&=-3 \text{Li}_{2,1,1}(-z,1,1)-\text{Li}_{2,1}(-z,1) \log (1+z),\\ \text{Li}_{1,1,2}(-z,1,1)&=\frac{1}{2} \left( \text{Li}_2(-z) \log ^2(1+z)+4 \text{Li}_{2,1}(-z,1) \log (1+z)+6 \text{Li}_{2,1,1}(-z,1,1)\right),\\ \text{Li}_{1,3}(z,1)&=-\text{Li}_{2,2}(z,1)-2 \text{Li}_{3,1}(z,1)-\text{Li}_{3}(z) \log (1-z),\\ \text{Li}_{1,3}(-z,1)&=-\text{Li}_{2,2}(-z,1)-2 \text{Li}_{3,1}(-z,1)-\text{Li}_{3}(-z) \log (1+z),\\ \text{Li}_{1,2,1}(z,-1,1)&=-\text{Li}_{2,1,1}(-z,-1,-1)-\text{Li}_{2,1,1}(-z,1,-1)-\text{Li}_{2,1,1}(z,-1,1)-\text{Li}_{2,1}(-z,1) \log (1-z),\\ \text{Li}_{1,2,1}(-z,1,-1)&=-\text{Li}_{2,1,1}(-z,-1,-1)-2 \text{Li}_{2,1,1}(-z,1,-1)-\text{Li}_{2,1}(-z,-1) \log (1+z),\\ \text{Li}_{1,2,1}(-z,-1,-1)&=-\text{Li}_{2,1,1}(-z,-1,-1)-2 \text{Li}_{2,1,1}(z,-1,1)-\text{Li}_{2,1}(z,-1) \log (1+z),\\ \text{Li}_{1,1,2}(z,-1,1)&=\text{Li}_{2,1,1}(-z,-1,-1)+2 \text{Li}_{2,1,1}(-z,1,-1)+2 \text{Li}_{2,1}(-z,1) \log (1-z)\\
&-\text{Li}_{1,1}(-z,-1) \text{Li}_2(-z)+\log (1-z) \log (1+z) \text{Li}_2(-z),\\ \text{Li}_{1,1,2}(-z,1,-1)&=\frac{1}{2} \Bigl( \text{Li}_2(z) \log ^2(1+z)+2 \text{Li}_{2,1}(-z,-1) \log (1+z)+2 \text{Li}_{2,1}(z,-1) \log (1+z)\\
&+2 \text{Li}_{2,1,1}(-z,-1,-1)+2 \text{Li}_{2,1,1}(-z,1,-1)+2 \text{Li}_{2,1,1}(z,-1,1)\Bigr),\\ \text{Li}_{1,1,2}(-z,-1,-1)&=\text{Li}_{2,1,1}(-z,-1,-1)+2 \text{Li}_{2,1,1}(z,-1,1)+\text{Li}_{2,1}(-z,-1) \log (1+z)\\
&+\text{Li}_{2,1}(z,-1) \log (1+z)+\text{Li}_{1,1}(-z,-1) \text{Li}_2(-z),\\ \text{Li}_{1,2,1}(-z,-1,1)&=-\text{Li}_{2,1,1}(-z,-1,1)-\text{Li}_{2,1,1}(z,-1,-1)-\text{Li}_{2,1,1}(z,1,-1)-\text{Li}_{2,1}(z,1) \log (1+z),\\ \text{Li}_{1,2,1}(z,1,-1)&=-\text{Li}_{2,1,1}(z,-1,-1)-2 \text{Li}_{2,1,1}(z,1,-1)-\text{Li}_{2,1}(z,-1) \log (1-z),\\ \text{Li}_{1,2,1}(z,-1,-1)&=-2 \text{Li}_{2,1,1}(-z,-1,1)-\text{Li}_{2,1,1}(z,-1,-1)-\text{Li}_{2,1}(-z,-1) \log (1-z),\\ \text{Li}_{1,1,2}(-z,-1,1)&=\text{Li}_{2,1,1}(z,-1,-1)+2 \text{Li}_{2,1,1}(z,1,-1)+2 \text{Li}_{2,1}(z,1) \log (1+z)+\text{Li}_{1,1}(-z,-1) \text{Li}_2(z),\\ \text{Li}_{1,1,2}(z,1,-1)&=\frac{1}{2} \Bigl( \text{Li}_2(-z) \log ^2(1-z)+2 \text{Li}_{2,1}(-z,-1) \log (1-z)+2 \text{Li}_{2,1}(z,-1) \log (1-z)\\
&+2 \text{Li}_{2,1,1}(-z,-1,1)+2 \text{Li}_{2,1,1}(z,-1,-1)+2 \text{Li}_{2,1,1}(z,1,-1)\Bigr),\\ \text{Li}_{1,1,2}(z,-1,-1)&=2 \text{Li}_{2,1,1}(-z,-1,1)+\text{Li}_{2,1,1}(z,-1,-1)+\text{Li}_{2,1}(-z,-1) \log (1-z)\\
&+\text{Li}_{2,1}(z,-1) \log (1-z)+[\log (1-z) \log (1+z)-\text{Li}_{1,1}(-z,-1)] \text{Li}_2(z),\\ \text{Li}_{1,3}(-z,-1)&=-\text{Li}_{2,2}(-z,-1)-\text{Li}_{3,1}(-z,-1)-\text{Li}_{3,1}(z,-1)-\log (1+z) \text{Li}_3(z),\\ \text{Li}_{1,3}(z,-1)&=\text{Li}_{2,2}(-z,-1)+\text{Li}_{3,1}(-z,-1)+\text{Li}_{3,1}(z,-1)-\text{Li}_2(-z) \text{Li}_2(z)-\log (1-z) \text{Li}_3(-z),\\ \text{Li}_{2,2}(z,-1)&=-\text{Li}_{2,2}(-z,-1)-2 \text{Li}_{3,1}(-z,-1)-2 \text{Li}_{3,1}(z,-1)+\text{Li}_2(-z) \text{Li}_2(z).
\end{aligned}
\end{equation*}

We end the Appendix by listing some useful identities we used for the evaluation of these functions \cite{BIGOTTE2002271, Broadhurst:1996az}:
\begin{equation*}
\begin{aligned}
\text{Li}_{1,1}(-1,-1)&=-\frac{\pi^2}{12}+\frac{1}{2}\log^2 2,\\
\text{Li}_{2,1}(-1,1)&=\frac{\zeta(3)}{8},\\
\text{Li}_{1,1,1}(-1,1,-1)&=\frac{\zeta(3)}{8}-\frac{1}{6}\log^3 2,\\
\text{Li}_{3,1}(-1,1)&=\frac{1}{48} \left(96 \text{Li}_4(1/2)+84 \zeta (3) \log 2-\pi ^4+4 \log ^4 2-4 \pi ^2 \log ^2 2\right),\\
\text{Li}_{3,1}(-1,-1)&=\frac{1}{2}\zeta(4)-2\left[\text{Li}_4(1/2)+\frac{\log^2 2}{24}(\log^2 2-\pi^2))\right],\\
\text{Li}_{1,1,1,1}(-1,-1,1,1)&=-\frac{3}{8}\zeta(2)^2-\frac{1}{4}\log^2 2\,\zeta(2)-\frac{1}{2}\text{Li}_{3,1}(-1,1)+\frac{7}{8}\log 2\, \zeta(3)+\frac{\log^4 2}{24}\\
&=-\text{Li}_4(1/2),\\
\text{Li}_{1,1,1,1}(-1,1,1,-1)&=-\frac{1}{40}\zeta(2)^2+\frac{1}{2}\text{Li}_{3,1}(-1,1)+\frac{\log^4 2}{24}.
\end{aligned}
\end{equation*}
In general, the following identities hold \cite{Broadhurst:1996az}
\begin{equation}
-\zeta(-1,-1,\{1\}_{l-2})=\text{Li}_l(1/2),
\end{equation}
\begin{equation}
-\zeta(-1,\{1\}_{l-1})=\frac{(-\log 2)^l}{l!}.
\end{equation}

\end{widetext}

\bibliography{refs}
\end{document}